# “Water” abundance at the surface of C-complex main-belt asteroids

Pierre Beck[1,2,], Jolantha Eschrig[1], Sandra Potin[1], Trygve Prestgard[1], Lydie Bonal[1], Eric Quirico[1] , Bernard Schmitt[1]

[1]Université Grenoble Alpes,CNRS, Institut de Planétologie et d’Astrophysique de Grenoble,

[2]Institut Universitaire de France, Paris, France

Abstract : Recently published space-based observations of main-belt asteroids with the AKARI telescope provide a full description of the 3-μm band, related to the presence of OH bearing minerals. Here, we use laboratory spectra of carbonaceous chondrites obtained under controlled atmosphere (CI,CM,CO,CV,CR Tagish Lake) to derive spectral metrics related to the water content in the samples. After testing several spectral metrics, we use a combination of band depth at 2.75 μm and 2.80 μm that shows a correlation with [$H_2O$] in the sample determined by TGA, though with a high uncertainty (4 wt. % $H_2O$). This relation is used to determine water content at the surface of large C-complex main-belt asteroids and discuss the origin of the variability found. On average C-complex Main-Belt Asteroids (MBA) have water contents of 4.5 wt.% (volume average, (1) Ceres excluded), significantly lower than average CM chondrites. The estimated water content for the most hydrated asteroids are lower than those of the most hydrated meteorites, a difference that could be attributed to space-weathering. An anti-correlation is also present between water content and overall spectral slope, which is opposite to expectation from laboratory simulations of space weathering on dark carbonaceous chondrites. This suggests that part of the variability in the surface hydration among the different C-complex asteroids is not due to space-weathering, but to the composition of surface material. When applied to Ceres, the hygrometer presented in this work enables us to estimate that at least 1.22 wt. % of the hydrogen is present in the form of organics. This richness in organics strengthens the connection between Ceres and cometary materials.

1. Introduction

The Solar System hosts a population of small objects but with large scientific value, the so-called small bodies (a denomination regrouping asteroids, comets, centaurs and trans-neptunian objects). Despite all these objects being small, they show a diversity of orbits, colors, densities or Vis-NIR spectra (Masiero et al. 2011, Vernazza and Beck, 2017, DeMeo and Carry, 2013, 2014 ; Carry, 2012). Among small bodies, main-belt asteroids (MBAs) are of particular interest since they are expected to be a major source of meteorites (Morbidelli and Gladman, 1998). Conversely studying meteorites may help to understand the nature of the main-asteroid belt population, and overall is a gateway to understanding the small bodies population as a whole. Among main-belt asteroids, the so-called S spectral-type have been linked successfully to ordinary chondrites (Nakamura et al., 2011), a family of meteorites that dominates the flux of meteorites since at least a few million years (Drouart et al., 2019). About half of the main-belt is composed of so-called C-complex asteroids (DeMeo and Carry, 2013), which have been connected to the less frequent carbonaceous chondrites group. More details on the link between the two groups will be provided by the two C-complex asteroid sample-return missions, Hayabusa2 (JAXA) and OSIRIS-REx (NASA).

While the S-complex asteroid / ordinary chondrites connection seems to be well established, the relation between carbonaceous chondrites and C-complex are more blurred. One of the key difficulties is the lack of strong absorption in the Vis-NIR for C-complex, with the exception of the 0.7 and 0.9 µm absorption bands, that have typical band depths of 5 % (Fornasier et al., 1999) and are attributed to $Fe^{2+}$-$Fe^{3+}$ bearing hydrated minerals. Many C-complex are devoid of these bands and their nature has been discussed whether as geologically processed (through thermal metamorphism, Hiroi et al., 1996) or as primitive (Interplanetary Dust Particles (IDP) rich, Vernazza et al., 2015). Because of their connection to carbonaceous chondrites, and the detection of signatures of hydrated minerals on some of them, C-complex asteroids are expected to be $H_2O$-rich. Carbonaceous chondrites do also show evidence of the presence of hydrated minerals in varying amounts depending on the carbonaceous chondrite class.

While absorption bands are faint on C-complex asteroids in the Vis-NIR, deep bands can be found in the 3-µm region where fundamental, harmonic and combination modes of –OH occur. These bands have been used to discuss the relations between carbonaceous chondrites and C-complex asteroids (Lebofsky et al., 1981, Hiroi et al.; 1996, Miyamoto and Zolensky, 1994, Sato et aL, 1997, Osawa et al. 2005, Bates et al., 2020, Takir et al. 2013, Beck et al., 2010, 2018, Potin et al., 2019, 2020, Rivkin et al., 2003, 2015, 2019). The comparison between laboratory spectra and telescopic observation has several complexities related to i) the contamination of laboratory data by terrestrial water, ii) the atmospheric absorption that masks the 2.6-2.9 µm region in ground-based observations, where the maxima of –OH absorption are typically found, iii) observation geometry effects, when comparing disk integrated observations to laboratory measurements.

Here, the approach is in essence similar to Rivkin et al. (2003) but we build on significant progress in these issues. First, reflectance of chondrite meteorites under asteroidal conditions have been obtained for about 40 samples (Potin et al., 2020b; Eschrig et al., 2020), and the effects of observation geometry on the 3-µm band have been recently quantified (Potin et al., 2019). Second, recently published space-based observations by the Akari telescope offer a full description of the 3-µm feature for several tens of Main-belt asteroids. This new laboratory dataset enables us to define metrics for remote quantification of water that are applied to derive the amount of water present at the surface of hydrated MBAs.

## 2. Methods

### 2.1 Reflectance datasets

The reflectance spectra used in this work have been published in previous articles. All these spectra were acquired with the same setup, the SHADOWS instrument (Potin et al., 2018), coupled to the MIRAGE environmental chamber. Unlike most instruments used to study reflectance in the 3-μm region, the SHADOWS system is not based on FTIR but uses a monochromatic light source and an InSb detector (Potin et al., 2018). Also, the SHADOWS instrument measures bi-directional reflectance spectra unlike most-systems that are biconical. All spectra are accessible through the SSHADE database (www.shade.eu).

The spectra of the CI Orgueil, Tagish Lake and CM chondrites are taken from Potin et al. (2019) and Potin et al. (2020b). The spectra of CO and CV chondrites are taken from Eschrig et al. (2020). Spectra of the Paris meteorite were presented in Bonal et al. (2019). All samples were measured under vacuum at 80°-100°C, with the exception of a few CO chondrites and the Paris lithologies, that were measured during an early measurement campaign under vacuum and 25°C. Details are provided in Table 1.

### 2.2 Quantification of equivalent "water"

Within meteorites, "water" may occur in several forms. Firstly, water may be present as molecules, in the form of adsorbed water, mesoporous water, structural water such as in hydrated salts (sulfates for instance) or fluid inclusions. Secondly, "water" may occur as –OH groups in the form of hydroxylated silicates, or hydroxides (Fe or Mg). In the later case, this not really water but can be somehow considered as water from a geochemical perspective, since they result from the interaction of anhydrous minerals and water in the parent body, and that molecular water will be released upon thermal decomposition of these minerals.

A largely used method for water and -OH quantification, used in Earth Sciences on geological samples, is infrared spectroscopy. However, this approach is difficult to apply to carbonaceous chondrites that are a complex mixture of complex phases. The techniques applied to derive water abundance in carbonaceous chondrites include X-Ray Diffraction (XRD) to estimate the amount of phyllosilicates (Howard et al., 2009, 2011, King et al., 2015, 2017), pyrolysis-based mass spectrometry (Alexander et al., 2012, 2013), thermogravimetry (Lee et al. 2016, Garenne et al. 2014, King et al. 2015, Gilmour et al. 2019, Bonal et al. 2020) as well as Karl-Fisher titration. Each of these techniques has pros and cons. The water abundances used for the present work were derived from thermo-gravimetric analysis (TGA, Garenne et al. 2014, Bonal et al., 2020), on the same sample or same batch of sample that was used for reflectance spectroscopy. This is an important point since many carbonaceous chondrites are breccias (Bischoff et al., 2006), and different fragments of the same meteorites may show contrasted water contents (Beck et al., 2014b). One of the disadvantages of TGA is that, except when it is coupled with mass spectrometry or infrared, it is not compound specific. This means that from TGA data we do not know if the mass loss is due to water, or if other compounds contribute (i.e. carbonates and organics). Still, they probably only contribute in a minor way since the TGA data (Garenne et al., 2014) were found to show a good correlation with H content determined independently (Alexander et al., 2012, 2013), as well as with IR spectroscopy (Beck et al., 2014a, Eschrig et al., 2020). In this work we used TGA data obtained in earlier studies (Garenne et al., 2014; Bonal et al., 2020). The mass analyzed are typically 30 mg of samples for TGA measurements, and 30-50 mg for reflectance spectra. The $H_2O$ content is taken as the 200-770° mass loss as used in Garenne et al. (2014) for CI and CM and as the 200-900°C mass loss for CO, CV and CR as used in Bonal et al. (2020). Note that this definition is different from that of King et al. 2015 that suggested to use 200-800°C for CI phyllosilicate dehydration.

3. Remote quantification

### 3.1 Lessons from transmission IR spectroscopy

Because OH bonds have a high dipolar moment they are active in infrared spectroscopy. The bending mode of water has a fundamental vibration around 6 µm while the OH stretching modes have fundamental vibrations around 2.7 µm for water and sometimes at higher wavelength for X-OH (where X is a transition metal). Combination and harmonic modes can be found in the vis-NIR (1.05, 1.4, 1.9, 2.1-2.3) and around 3-µm. The so-called 3-µm band is a composite feature extending from 2.7 to about 4-µm with contributions from fundamental, harmonic and combination of stretching and bending vibration.

Transmission spectroscopy is sensitive to water related absorptions and can be used to quantify water abundance down to ppm levels (for example in tektite; Beran and Koeberl, 1997). In previous works, we studied water "speciation" and abundance of $OH/H_2O$ within carbonaceous chondrites (bulk sample: Beck et al., 2010, 2014, matrix: Bonal et al., 2016, 2020) and micrometeorites (Battandier et al., 2018). These works revealed that under ambient conditions adsorbed terrestrial water contributes significantly and contaminates the 3-µm region (Beck et al. 2010). They also showed a change of the position of the absorption maximum (Beck et al. 2010) for CM chondrites and a correlation between the intensity of the 3-µm band (normalized to the silicate feature intensity) and hydrogen content (Beck et al., 2014).

### 3.2 Complexities associated with reflectance

While transmission infrared is efficient to quantify water content, the use of the 3-µm band in reflectance is more challenging. Reflectance spectroscopy involves both absorption and reflection, and the resulting spectrum is controlled by composition (absorption properties) and physical properties (refractive index, grain sizes, porosity) and observation geometry. Detailed work by Milliken and Mustard (2005) has revealed some of the complexities in defining a unique water abundance calibration curves based on the 3-µm band.

To illustrate some of the difficulties, synthetic mixtures of an -OH bearing phase with a dark component were modeled using Hapke (2012) theory. A dark and spectrally flat component was mixed with a brighter component to which an –OH absorption feature was added in the form of an exponentially modified Gaussian (Potin et al., 2020a)(Fig. 1). Simulations were run so that the volume fraction of the –OH free and –OH bearing compounds were kept constant, implying that the amount of water in the mixture was constant. However, the relative grain size between the two constituents was changed. Calculations showed that the band depth could vary by a factor of ten, in the grain size range investigated (diameter ratio varying between 0.05 and 500, and the diameter of the bright grain being 10-µm). When the dark grains are large, the mixture is relatively bright and the dependence of band depth on the size of the dark grains is modest. When the size of the dark grains is smaller than the size of the bright grains, then the mixture is dark and the dependence of band depth to the size of the dark grains is high. Note that these calculations do not apply to grain sizes that are much smaller than wavelength and should follow a different regime of radiative transfer.

In this model case, the band depth alone is found to be a degenerated parameter, which will fail in retrieving water content: samples with the same water content have different band depth. Applying radiative transfer models that try to reproduce not only band depth but also absolute reflectance may help in this matter, since they can provide constraints on grain size and remove some of the degeneracy. However, they rely on a certain number of parameters (nature and intensity of the opposition effect, shape of the single particle phase function, optical constants,…) that may be difficult to constrain and often need an a priori hypothesis. Another approach that may be used is to analyze samples that can be seen as simulants of asteroids surfaces, namely meteorites.

### 3.3 Calibration curve

Using meteorites as simulant of asteroidal surfaces has shown great success in predicting the composition of S and V type asteroid (Nakamura et al. 2011; DeSanctis et al. 2012). Even if the meteorite is not exactly made of the same material as the surface of the asteroids, it may provide a good analogue in terms of grain size, texture and mineralogy, and can then release some of the degeneracies associated with reflectance spectroscopy.

Several metrics to quantify water abundance from reflectance spectra have been defined and investigated in the past, in particular in Milliken and Mustard (2005) and Pommerol and Schmitt (2008). Metrics specific to carbonaceous chondrites have been discussed in Rivkin et al. (2003), Kaplan et al. (2019), Garenne et al. (2016) and Beck et al. (2018). The specificity of the work discussed in the present article is to try to define metrics that are applicable to different classes of carbonaceous chondrites, and to use newly acquired spectra on CR, CO and CV chondrites. Here, we tested several of those metrics. We will discuss them in the next paragraph.

One of the difficulties in ground-based investigations of small bodies in the 3-µm region is the presence of water-vapor and ice in the earth-atmosphere that mask a significant portion of the absorption, including the absorption maximum for most phyllosilicates. A method that has been used to circumvent this problem is to use a combination of band depth at 2.9 and 3.2 µm (Sato al., 1997; Rivkin et al., 2002), outside of the Earth's atmospheric absorption. A good correlation between these two band depths was found previously (Sato et al., 1997, Takir et al., 2019) as well as in our dataset obtained on carbonaceous chondrites (Figure 2). Within this array the CM chondrites appear to define a line offset from that defined by CV, CO and CR chondrites, which can be explained by different shapes of the 3-µm band. The CM chondrite (as well as the CR1 GRO 95577) have a band with a maximum of absorption around 2.7-2.80 µm while CO, CV and CR tend to have an absorption maxima in the 2.85-3.0 µm range (Eschrig et al., 2020). Figure 3 presents the relation between carbonaceous chondrites water content and the band depth at 2.9 µm. This graph shows a strong dispersion in the dataset and that the different classes of carbonaceous chondrites show different relations between water-content and band depth at 2.9 µm. More specifically the slope of the band depth vs $H_2O$ content seems to be lower in the case of CM chondrites than in the case of CV-CO chondrites. This can be explained by the fact that OH may be present in the form of oxides in some of the samples (CV) rather than in phyllosilicates. These distinct behaviors may also be related to the fact that the petrographic constituents (Chondrules, matrix, metal, CAI) are not distributed in the same way between those various meteorite groups and are not of the same size (Krot et al., 2006). Post-accretion thermal history may also have played a role in shaping the reflectance spectra of these samples.

Because the energy of the absorption is spread among the whole band, band areas or integrated band depths are criteria often used in spectroscopy. In figure 4, we present the integrated 3-µm band depth (between 2.6 and 3.8 µm) as a function of the water content, and a behavior similar to the 2.90 µm band depth is seen. After testing several metrics, the sum of the 2.75 and 2.80 µm band depths was found to correlate best with water content (Figure 5). Still, the quality of the correlation is modest. Such a combination of band depth was already tested in previous studies (Garenne et al., 2016; Beck et al., 2018), with the idea to define a criteria that would be specific to CM chondrites phyllosilicate, that tend to have band maxima in the 2.75-2.80 µm range, at shorter wavelength than adsorbed water (2.85 µm) or water ice (3.1 µm). The fact that this criterion shows the best correlation with water content when considering carbonaceous chondrites in general appears somehow fortuitous given the difference in 3-µm band shape between CM, CR, CV and CO. The difference in band shape implies different mineral host of -OH groups, which a priori should have different molar absorption coefficient at a given wavelength. To explain this improved correlation, one may then invoke that molar absorption coefficient are about the same at 2.75 and 2.80 µm for all these samples, which could in principle be tested but not simply (this may need transmission measurement of samples of well controlled thicknesses). The other possibility is that because CO and CV chondrites are brighter than CM chondrites, for a given water content, band depth are smaller for CM chondrites than CO and CV, as found in our simulations (part 3.2 and figure 1). By using a wavelength which in the case of CV and CO chondrite is on the wing of the absorption band, this effect is somehow compensated.

Still overall, the correlation in Figure 5 is not great which may be due to errors on the measurements of water content. This error can be estimated by comparing the water content measured with TGA to that measured by mass spectrometry, and should be of the order of 2 wt.% $H_2O$. This is not enough to explain the spread observed in figure 5, which is rather attributed to a true spectral diversity, in relation with grain sizes, mineralogy, which themselves are related to the often-complex accretion and metamorphic history of these samples. Still, this trend enables us to remotely infer water content for samples expected to be “similar” to carbonaceous chondrites, though with large error. Errors on the retrieved amount of water content were estimated by calculating confidence bands to the linear regression of the whole dataset presented in figure 5. This resulted in typical errors of +/- 4 wt.% for a 90 percent confidence.

## 3.4 Temperature, grain size and observation geometry effects

Some limitations apply to the “hygrometer” proposed above, which was defined using powders (typical grain size 50-150 µm, Garenne et al., 2016) measured under standard observation geometry, under vacuum and 80C.

### 3.4.1 Effect of temperature

Temperature has an effect on the 0.4-2.5 μm reflectance spectra of ordinary chondrites (Hinrich and Lucey, 2002), and corrections need to be applied to determine remotely the composition of S-type asteroids (Dunn et al. 2013, Burbine et al. 2002). Less is known in the case of carbonaceous chondrites, and the change of the 3-μm band of carbonaceous chondrites when exposed to low-temperature has not been studied to our knowledge. While molecular water in the form of ice or hydrated minerals has a 3-μm band that will change with temperature because of slight re-organization in the crystal structures, metal-OH in phyllosilicates are not expected to show a strong variation of their absorption when exposed at main-belt asteroid temperature as was observed for the Ceres simulants measured under Ceres-like temperature (Galliano et al., 2020).

#### 3.4.2 Powder vs rock

Because of collisions occurring on their surfaces, a layer of fine powder, analogue to the lunar regolith is likely to cover large main-belt asteroids. Typical grain size determined for lunar soils is of 0.1 mm (Carrier, 1973). On the contrary small near-earth asteroids, and to some extent possibly small main-belt asteroids, appear to be rubble-piles, and therefore may have a surface covered by rocks rather than by fine particulate materials, as observed for asteroid Bennu (DellaGiustina et al., 2019). This difference in texture is reflected in the difference in thermal inertia between small and large asteroids (Delbo et al., 2015). In order to investigate the difference between the hydration signatures of powder and rocks, fragments of the CM fall Aguas Zarcas were measured before and after grinding (Fig. 6). Results showed systematic differences between rock and powder including, a change of slope from blue to red, more pronounced 0.7 and 0.9 μm feature in the case of the powder, and a deeper 3-μm band for powders than for rocks. Because the measurements were not obtained under vacuum, the exact shape of the 3-μm band cannot be compared to the rest of the dataset. Still, when considering absorption depth at 2.75 μm (where adsorbed water contribution should be more limited) the band depth is typically 35 to 60 % larger in the case of powder than rocks. This analysis suggests that applying the calibration defined in figure 5 to an asteroid covered by rocks rather than regolith will result in an underestimation of the water content. Note also that differences in the spectral signatures of CM powders and thin section were also observed by Hanna et al. (2020).

#### 3.4.3 Observation geometry

Reflectance spectra of meteorites are often measured in the laboratory under "standard" geometry, i.e. with a phase angle of 30° and a usually low-incidence angle. Ground-based observations of main-belt asteroids are obtained at phases ranging from 0 to 30° and are integrated over a whole hemisphere. As a consequence, the phase angle is usually different between laboratory data and ground-based observations but incidence and emergence as well. Meteorite powders and rocks do not behave as lambertian surfaces (Gradie et al., 1980, Capaccioni et al., 1986, Beck et al., 2012, Potin et al., 2019), which can have an impact on band depth. Typically, if we exclude extreme observation geometries (incidence and emergence above 50°) the variation of the 2.75-μm band depth of a powdered CM chondrites was found to be of the order of 20% within the range of geometry studied by Potin et al. (2019). While it may be worth modeling the effect of observation geometry on the band depth for an asteroid with topography, it may not impact the remote estimate of water content too significantly.

4. Application to C-complex MBAs

## 4.1 The amount of water

The new combined band depth parameter we defined was applied to space-based observations of C-complex main-belt asteroids from Usui et al. (2018) and to orbit-based observation of Ceres (DeSanctis et al., 2015; Marchi et al., 2018) (Table 2). The shape of the 3-µm band varies among the studied objects (see Table 1). For most C-complex the band shape resembles that of CM chondrites (sharp-type, Usui et al., 2018; Potin et al., 2020) but for a few asteroids the band shape is different. Two objects (Ceres, Hygiea) have a small absorption at 3.06 µm interpreted by ammonium ions in phyllosilicates (King et al., 1992), and a few asteroids have a broader feature with a maxima at 3.1 µm that has been interpreted by a thin layer of water-ice (Rivkin and Emery, 2010; Campins et al., 2010), goethite (Beck et al., 2011), or ammonium-salts (Poch et al., 2020). Given the fact that our criteria uses a lower wavelength (2.75-2.80 µm) than where ammonium ion or thin layer of water-ice absorb, it should not be too much perturbated by the presence of these compounds, and will enable to retrieve the amount of "water" present in a same form as in carbonaceous chondrites. Water ice will not be accounted for with our criteria.

The band depth at 2.75 and 2.80 microns were calculated (with a continuum at 2.6 µm), summed and converted into water content using the regression line in figure 5. The values found are between 0 and 10.3 wt. % for the studied dataset. Water contents for some of the asteroids investigated with Akari (Usui et al., 2018) were also determined by Rivkin et al. (2003). The values derived in the present work are slightly lower than those found in Rivkin et al. (2003) with the exception of 2-Pallas for which the estimate in the present work is higher than their value. Note that within the limit of the calibration errors presented earlier the estimates from this work and those of Rivkin et al. (2003) agree.

In figure 7, the water content derived for C-complex is plotted against the asteroid diameter. This graph does not reveal obvious relations between the derived amount of water and the size of the asteroid. Comparison of the derived water content also reveals that at present there are no large asteroids that are a match to the water-rich CI chondrite Orgueil. This is somewhat paradoxal since, among meteorites, CI chondrites have the chemical composition that is closest to the solar photosphere, suggesting that CI-precursor like material should have been the dominant dust within the solar protoplanetary disk. Two propositions can explain this paradox: whether CI-precursor like material did not accrete onto small bodies, or the small bodies that accreted CI-precursor like material only rarely experienced aqueous alteration.

This graph also confirms the observations previously noted by Rivkin et al. (2003) and Hiroi et al. (1996) that C-complex asteroids appear less hydrated than the most hydrated carbonaceous chondrites. As studied by several authors, aqueous alteration has been found to be variable among carbonaceous chondrites, and the definition of an extent of aqueous alteration has been debated (Rubin et al., 2007, Alexander et al., 2013, Howard et al., 2009, Beck et al., 2014). Generally, the total amount of $OH/H_2O$ present in carbonaceous chondrites appear to vary in the order CI > CM >CV-CO-CR. Some exceptions to this trend have been found. Some CO or CR can show unusually large amount of water (the CR1 GRO95577 or the CO MIL 07687). Also, an important fraction of CM chondrites has been classified as thermally metamorphosed, and contain lower amount of water than “normal” CM (Nakamura et al., 2005, Alexander et al., 2012, Garenne et al., 2014, Quirico et al., 2018).

When comparing the water content derived from B&C objects observed by Usui et al. (2018), it is found that their water content are more similar to those derived for CO-CV-CR, and lower that non-metamorphosed CM. However, the position of the maxima of absorption is generally at shorter wavelengths than that measured for CO-CV-CR (Eschrig et al., 2020).

4.2 The hydration-dehydration trend

There are three hypotheses that have been formulated to explain the distribution of water observed for C-complex: space weathering, thermal metamorphism, and the presence of unlithified IDP-related objects.

Space weathering can be described as the ensemble of processes acting at the surface of an airless body (Pieters and Noble, 2016). These processes include irradiation by the solar wind and galactic cosmic rays, as well as impacts by micrometeorites. Space weathering has been shown to significantly modify the optical properties of ordinary chondrites and S-type asteroids (Chapman, 2004). A number of recent studies have investigated the impact of both processes on carbonaceous chondrites-like surfaces by ion or laser irradiation to simulate solar wind effects and micrometeorite impacts respectively (Matsuoka et al., 2020; Lantz et al., 2017). Both processes were found to induce a progressive but moderate bluing of the spectra as well as a progressive decrease in the 3-μm band, with increasing irradiation dose or laser intensity. In order to search for these signatures of space weathering on large C-complex asteroids, the intensity of hydration is plotted against the spectral slope (R2.45.R0.55) in figure 8. This graph reveals however an increase of slope with decreasing hydration, which is the opposite of the trend observed in space weathering experiments. This observation does not mean that space weathering is not active on C-complex. Indeed, the difference between CM chondrites and the most altered C-complex asteroids observed in figure 8 could be explained by space weathering. However, the observed trend suggests that the variability among C-complex main-belt asteroids is not due to different intensities of space weathering but rather to variable compositions.

Thermal metamorphism has been proposed as a mechanism to explain the general depletion of C-complex asteroids in water when compared to CM chondrites (Hiroi et al., 1996). Thermal metamorphism may have occurred at the surface of small bodies as a consequence of radioactive decay in the first tens of millions of years of the Solar System, or as impact-induced heating over a longer period of time. The identification of heated CM chondrites supports the idea that the surface of C–types may have been thermally dehydrated. This view if supported by observations of the NEA Ryugu by the Hayabusa-2 spacecraft (Kitazato et al., 2019). In the case of MBAs, several challenges to this hypothesis have been discussed in Vernazza et al. (2015). They include observation in the 10-μm region, the overall low-density of C-complex and the proportion of heated CM among CM chondrites. The favored hypothesis in Vernazza et al. (2015), discussed also in Rivkin et al. (2019), is rather that the non-hydrated C-complex are primitive bodies that rather deliver samples to Earth in the form of interplanetary dust particles. Because heated CMs tend to have noticeable water content and 3-μm band depths, the presence of water-poor C-complex would tend to favor the hypothesis of Vernazza et al. (2015).

The reader should also keep in mind that present-day delivery of carbonaceous chondrites is unlikely to be samples directly ejected from the large C-complex asteroids discussed here. The

A last possibility, that has not been formulated previously, is that the amount of terrestrial water present in carbonaceous chondrites is underestimated, and subsequently that the amount of asteroidal water in carbonaceous chondrites is overestimated. Recent analysis of Vacher et al. (2020) may support this hypothesis. Indeed, the amount of hydrogen derived for CM samples exposed to 120°C and vacuum for 48h were significantly lower than those estimated by Alexander et al. (2012). This is of particular interest for Orgueil, the sample that shows the largest change in the shape of the 3-µm band when exposed to vacuum and moderate heating (Beck et al., 2014; Potin et al., 2020). The spectral slope of powders from the fresh fall Aguas Zarcas were not found to be significantly bluer that other CM chondrites (1.32;1.38;1.57). This means that if terrestrial oxidation occurs and modifies mineralogy and water content of CM chondrites powders, this may be a quasi-immediate effect. Samples from Bennu and Ryugu will certainly be beneficial in understanding possible issues with contamination by terrestrial water, if they can remain unexposed to terrestrial water before reflectance spectra measurements and water abundance estimates.

### 4.3 The Ceres case

Ceres is the largest main-belt asteroid and is now considered as a past ocean world. As observed from orbit and by the Dawn mission, Ceres is covered by dark hydrated material (Lebofksy et al., 1981; DeSanctis et al., 2015). The 3-µm region of Ceres is distinct from most C-complex asteroids, since signatures of ammonia have been observed, suggesting an outer Solar System signature (DeSanctis et al., 2015) and a connection to comets (Poch et al., 2020). Only a few asteroids share this peculiar signature (Takir et al., 2013; Rivkin et al., 2019).

In addition to a detailed mapping of the surface of Ceres in the Vis and IR, the Dawn mission also remotely determined the amount of hydrogen present in the top first meter of Ceres (Prettyman et al., 2016). These measurements revealed the presence of "permafrost" in the high latitude regions of Ceres and a typical hydrogen content of 1.9 wt.% (±0.2) in equatorial regions (Prettyman et al. 2016, Marchi et al., 2018).

By applying the calibration of water content presented here, it is possible to estimate how much water is present at the surface of Ceres, and therefore how much hydrogen detected by GRaND is related to -OH in phyllosilicates. Using typical reflectance spectra of Ceres in the 3-µm region (Marchi et al., 2018) the water content derived for Ceres equatorial region is 5.2 wt. % ( ±4), which converts into a [H wt.%] of 0.58 wt. % (±0.44). This means that at least 1.32 wt. % (±0.44) of hydrogen in the equatorial regions of Ceres is present in other forms than hydrated minerals, the candidates being $NH_4^+$ and H-bearing organics as observed by the VIR instrument (DeSanctis et al., 2015). In order to estimate how much hydrogen may be present in the form of $NH_4$+, we used the experimental reproduction of Ceres spectra presented in Galiano et al., (2020). The best match to Ceres average spectra was obtained with a mixture using 28 wt. % of $NH_4^+$-montmorillonite. Within $NH_4^+$-montmorillonite, hydrogen present in $NH_4^+$ is estimated to be around 0.3 wt. % (Pironon et al., 2003), which converts into an estimated abundance of hydrogen contained in ammonium ions of around 0.1 wt. % of the equatorial material of Ceres. The hydrogen content remaining for organic compounds is therefore 1.22 wt.% (±0.44). If the organic compounds are similar to CM chondrite insoluble organic matter (IOM) with an H/C close to 0.65 this converts into a carbon content of 22.5 wt. % (±8.1). This is slightly higher but in reasonable agreement with one of the models proposed for Ceres surface composition (20.2 wt. %, Marchi et al., 2019). Such very high values have also been proposed by Kaplan et al., 2018 but for the organic hotspots. If the organic carbon is present in the form of less refractory organics, similar to the soluble organic compounds encountered in meteorites then its H/C may be much higher and the carbon content on the surface of Ceres lower (for instance, with H/C =1.55 for methanol extract in Schmitt-Kopplin et al., 2010, the carbon content would be around 9.4 wt. %(±3.4).

It is important to stress that these estimates of organic content at the surface of Ceres (as well as the ones of Marchi et al., 2018) rely on several hypothesis, one of the strongest being that the surface of Ceres is vertically homogeneous within the depth probed by both infrared and neutron. This hypothesis is reasonable since it is probable that similarly to the Moon surface, gardening induces some level of mixing in the top meter at geological timescales. Under the hypothesis that Ceres' organics are similar to meteorite IOM or soluble organic matter (SOM), the amount of carbon obtained here is much higher than the C-content of bulk carbonaceous chondrites (at. C/Si about 0.8), but in the range of values found for IDPs (at. C/Si around 2) or similar to values derived for comet 67P (at. C/Si=5.5) (Bardyn et al., 2017, Herique et al., 2018). In addition to the identification of ammonium ion on both Ceres and comets (Poch et al., 2020; Altwegg et al. 2020), the richness in carbon strengthens the connection between Ceres and cometary materials.

### 4.4 Total amount of water in the B&C reservoir

The volume average for the studied asteroids is 5.0 wt.% (4.5 wt. %, excluding Ceres). Under the hypothesis that this value is representative of the bulk material constituting the C–type population of asteroids, we can use the mass of this population by DeMeo and Carry (2013) to estimate the amount of water they represent (this calculation neglects the possible presence of ice in their interior). This is certainly a strong assumption but at the moment there is little evidence for internal heterogeneity for C-complex (at least for Ch-Cgh). First a Vis-NIR survey of Vernazza et al. (2016) revealed that Cgh-Ch asteroids of different sizes show very little spectral diversity with the exception of a change of slope that can be attributed to a change of grain size. Second in the case of (41) Daphne, one of the few C-complex asteroids with constrain on the internal structure, observation of its shape and determination of its density do not support differentiation (Carry et al., 2019).

The assumption that the [$H_2O$] value derived is representative of the object as a whole implies that exogeneous process do not modify the water content. If we make the hypothesis that the C-complex richest in [$H_2O$] is a space-weathered version of the most altered CM, this means that we underestimate water content of C-complex by 20-30 %. Also, impact may produce dehydration of surface material, but at least in the case of Ceres the 2.7 μm band depth shows limited variability across the surface, and the variability does not seem to be related to cratering (Ammannito et al., 2017). Also note that when making the estimate of total amount of water in the C-type reservoir, we neglect the present of water-ice in theire interior. Water-ice is suspected to be a major constituent of Ceres interior, and may also be present in the interior of C-type asteroids in general. However, it is difficult at the moment to estimate how much water ice (and possibly also other type of ices) is present within C-type.

With the above-mentioned limitations in mind, the value obtained for the amount of water in the C-complex reservoir is of 7.1 x $10^{19}$ kg, i.e. 3/4 of the mass of Enceladus. As a whole, the C-complex reservoir only represents 0.5 % of the total amount of water present on Earth (Marty, 2012). Note that the fact that this amount of water derived is much lower than that found on Earth does not preclude the fact that C-complex asteroids may have been a major source of volatiles in the inner Solar System (Alexander et al., 2012). Main-belt C-complex asteroids are thought to be implanted from a more distant source region (beyond Jupiter), as a consequence of the dynamical evolution of the Solar System (Raymond and Izodoro, 2017; Walsh et al., 2012). In other words, the initial mass of the C-complex reservoir was probably several orders of magnitude higher than what can be weighted today in the main-belt.

## 5. Conclusions

In this work we discuss an approach to quantify the equivalent water abundance at the surface of main-belt asteroids from space-based observations of the 3-µm absorption band in comparison to laboratory measurements on carbonaceous chondrites obtained under a controlled atmosphere. From this work, the following conclusions can be drawn.

- When using band depths above 2.9 µm (i.e. outside of the atmospheric vapor absorption), there is no unique relation between band depth and water content that is valid for all classes of carbonaceous chondrites studied here (CI, CM, CO, CV ,CR).

- A metric is defined that appears to work for all classes of carbonaceous chondrites studied, using a combination of band depths at 2.75 and 2.80 µm. A significant error on the estimate exists (typically 4 wt. %), that is inherent to the complexity of radiative transfer in a complex media like carbonaceous chondrites.

- This metric is used to infer the water content on large B & C types, which were investigated in the 3-µm region by Usui et al. (2018) using the Akari space telescope. In the case of asteroids which were also studied in Rivkin et al. (2003) using ground-based observations, the derived values are in agreement within the error. The derived values for water content range between 0 and 11.5 wt. % (+/- 4 wt.%, 90 % confidence). The average value for the 20 objects investigated is 5.0 wt % (4.5 wt. % if Ceres is excluded). If this value is taken as a representative of the whole population, the C-complex ”water” reservoir is of 7.1x$10^{19}$ kg, which is about the mass of Enceladus or 0.05 Earth oceans.

- The water content calculated for the most hydrated asteroids are lower than those of the most hydrated meteorites including Orgueil. This difference may be attributed to space weathering or to an underestimated contribution of terrestrial water to the carbonaceous chondrites water budget.

- An anticorrelation is found between the water content and the vis-NIR slope for B & C type MBA. This is opposite to the expected space weathering trends and appears to reflect a continuous mixing between two styles of materials.

- Applying this water quantification tool to Ceres enables the estimation of the amount of hydrogen present in the form of organics, within the equatorial region. At least 1.22 wt. % of hydrogen is expected to be present in the form of organics. This richness in organics strengthens the connection between Ceres and cometary materials.

ACKNOWLEDGMENTS:
This work was funded by the European Research Council under the H2020 framework program/ERC grant agreement no. 771691 (Solarys). Additional support by the Progamme National de Planétologie and the Centre National d'Etude Spatiale is acknowledged. Comments by Ashley King and an anonymous reviewer greatly improved the manuscript. Vincent Jacques and Emmanuel Jehin are acknowledged for providing the Aguas Zarcas samples.

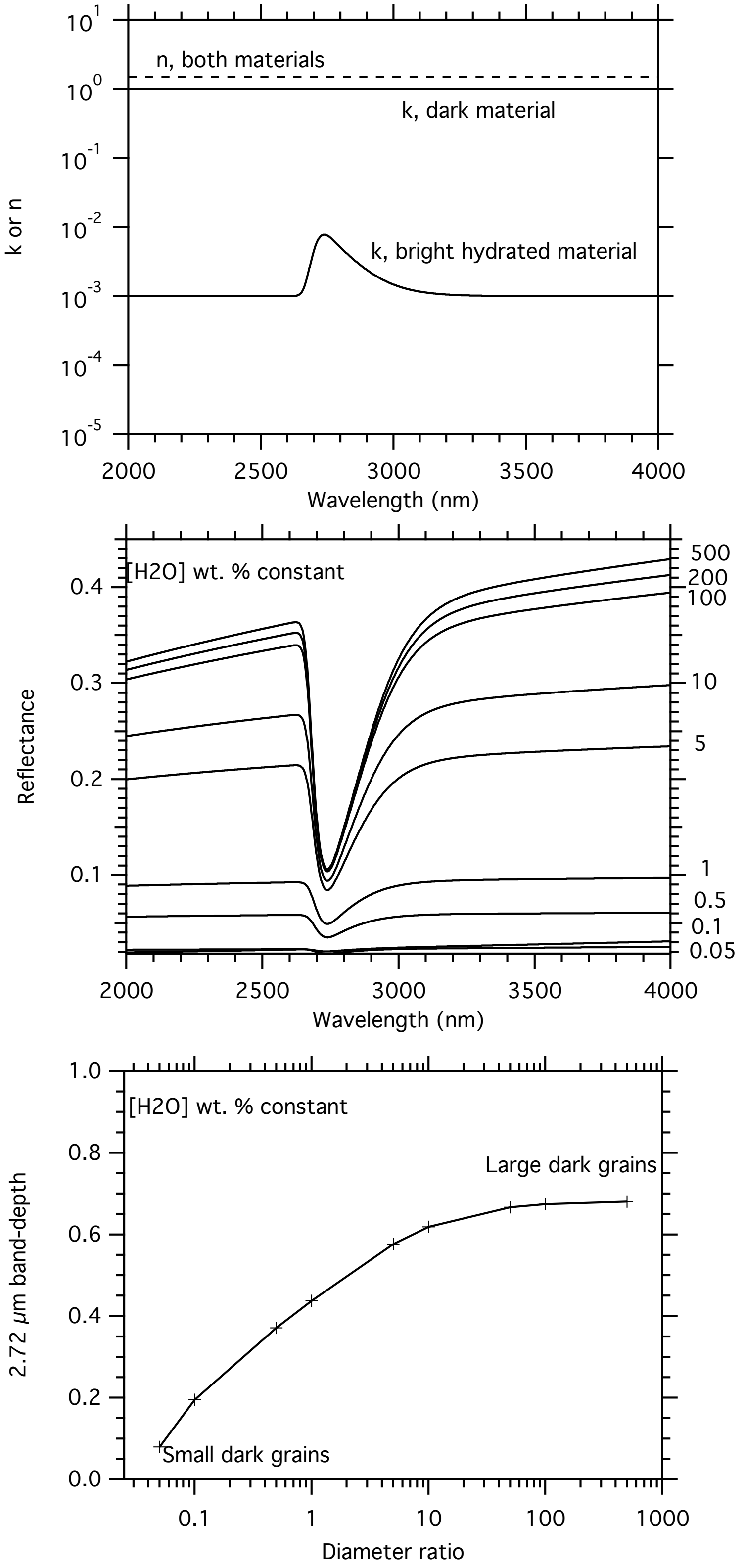

FIGURE 1: Mixing model of hydrated material with a dark component, using synthetic optical constants. Top) Optical constants used for these calculations. Middle) Modeled reflectance for changes in the grain size ratio. Bottom) 2.72 μm band depth calculated for the various mixtures. Note that the water content of the mixture is constant but the relative grain size of the two constituents is changed.

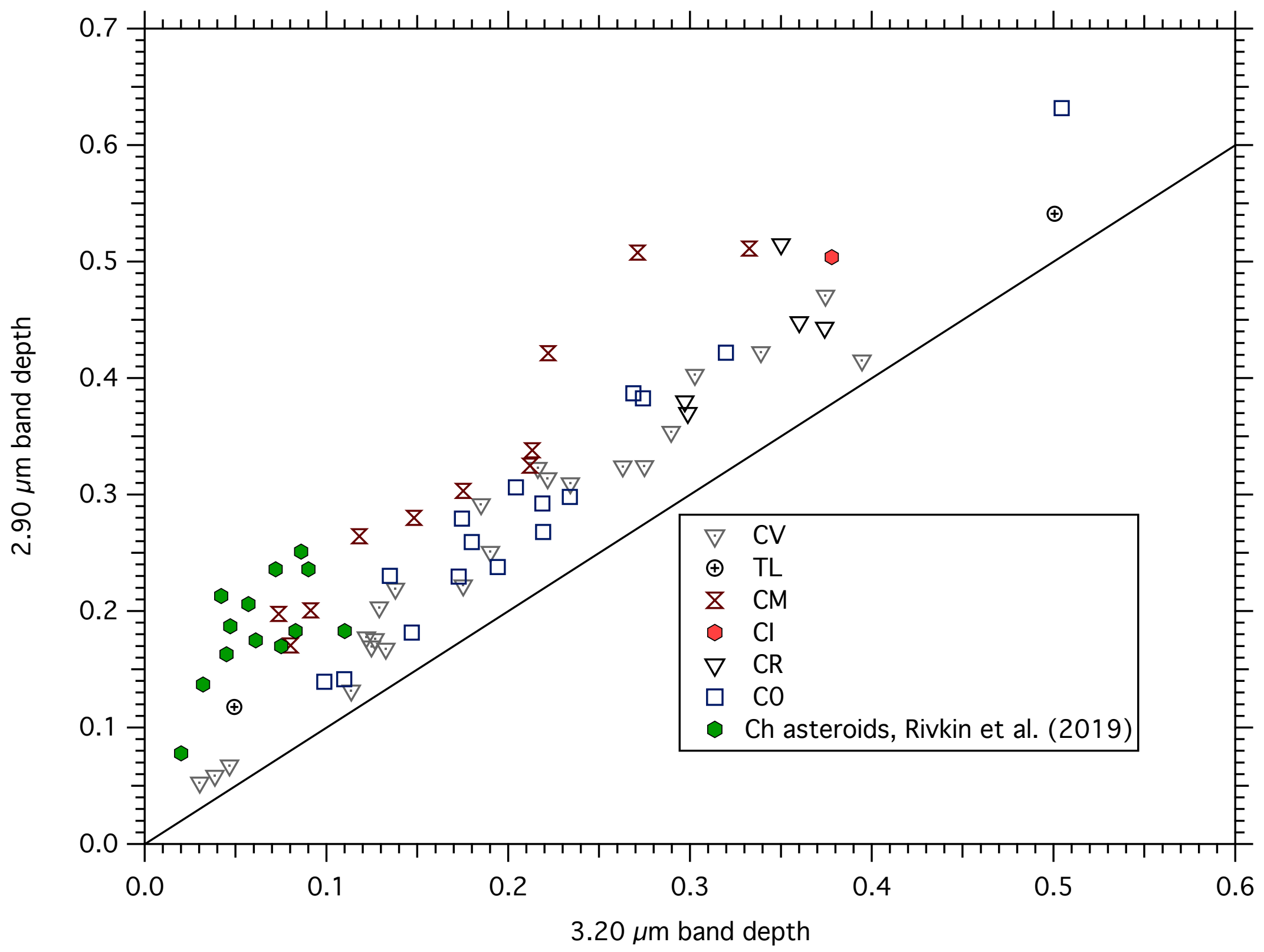

FIGURE 2: Band depth at 2.90 μm against band depth at 3.20 μm of meteorite powders of different classes measured under simulated asteroidal conditions. The black line is y=x. The two Tagish Lake measurements correspond to two different lithologies.

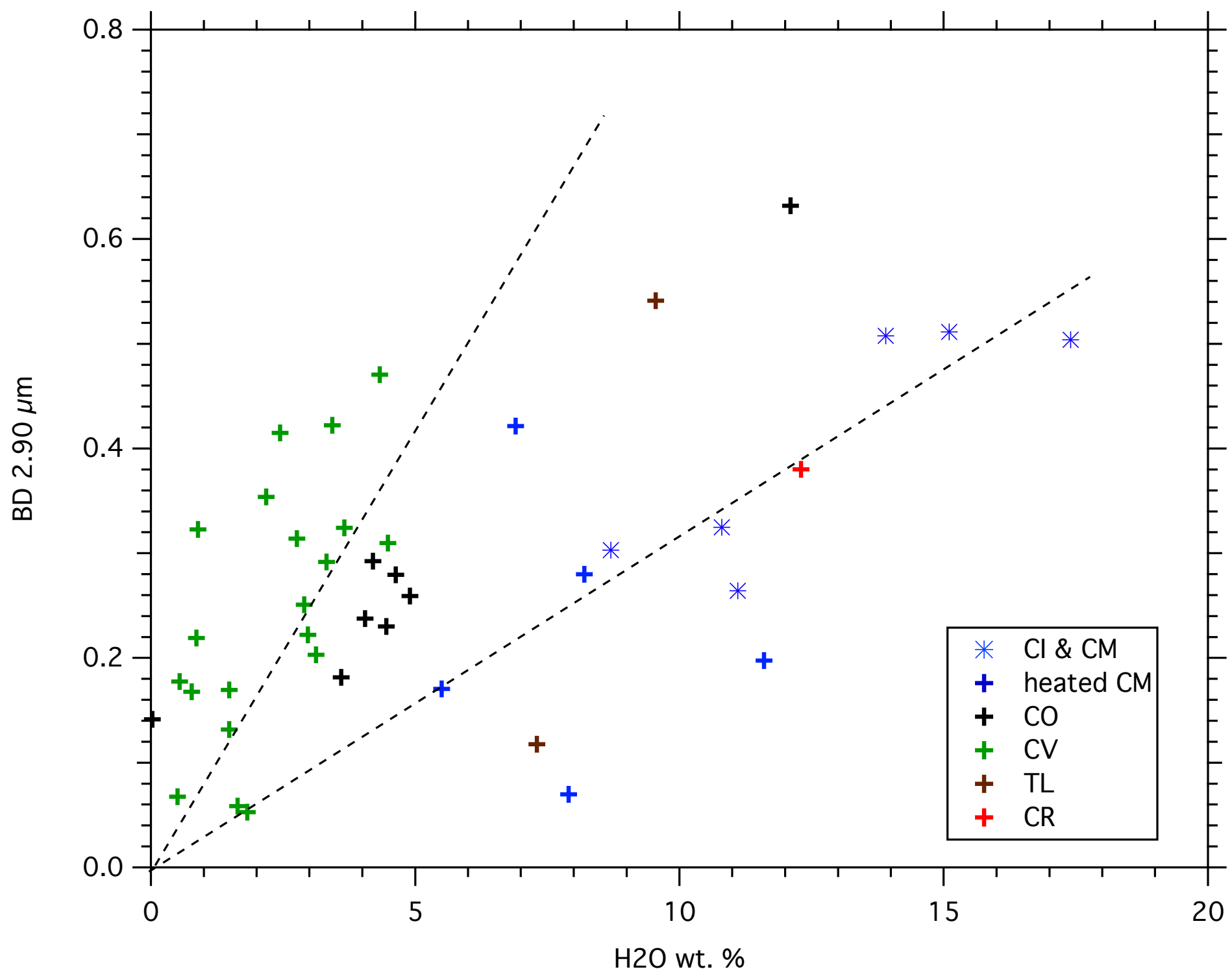

FIGURE 3: The 2.90 µm band depth as a function of water content (TGA analysis) for different classes of carbonaceous chondrites. Two distinct trends appear for CI & CM and CO-CV. A unique quantification scheme for carbonaceous chondrites is not possible.

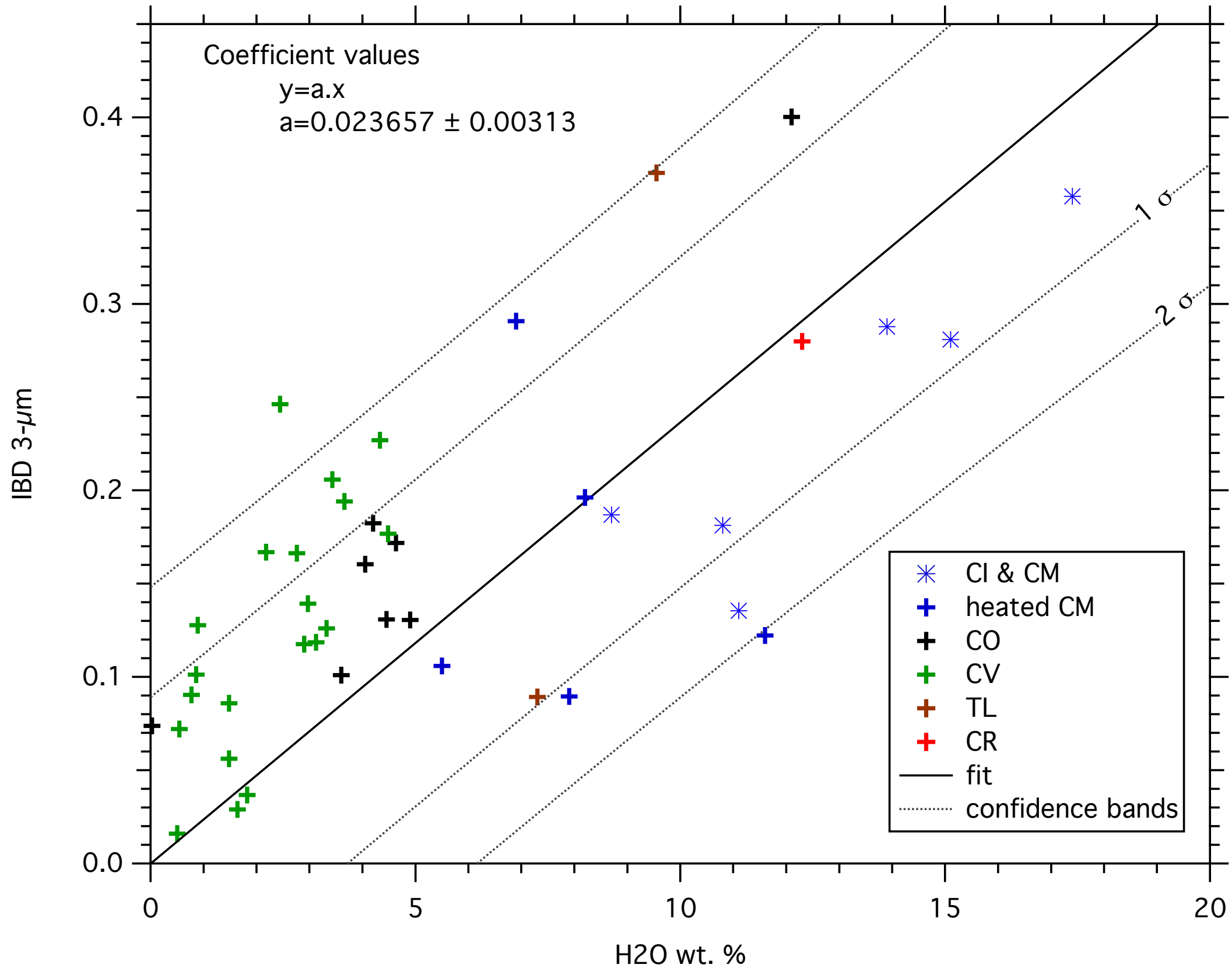

FIGURE 4: The integrated 3-μm band depth as a function of water contents (TGA analysis) for different classes of carbonaceous chondrites. Two distinct trends appear for CI & CM and CO-CV. A unique quantification scheme for carbonaceous chondrites is not possible.

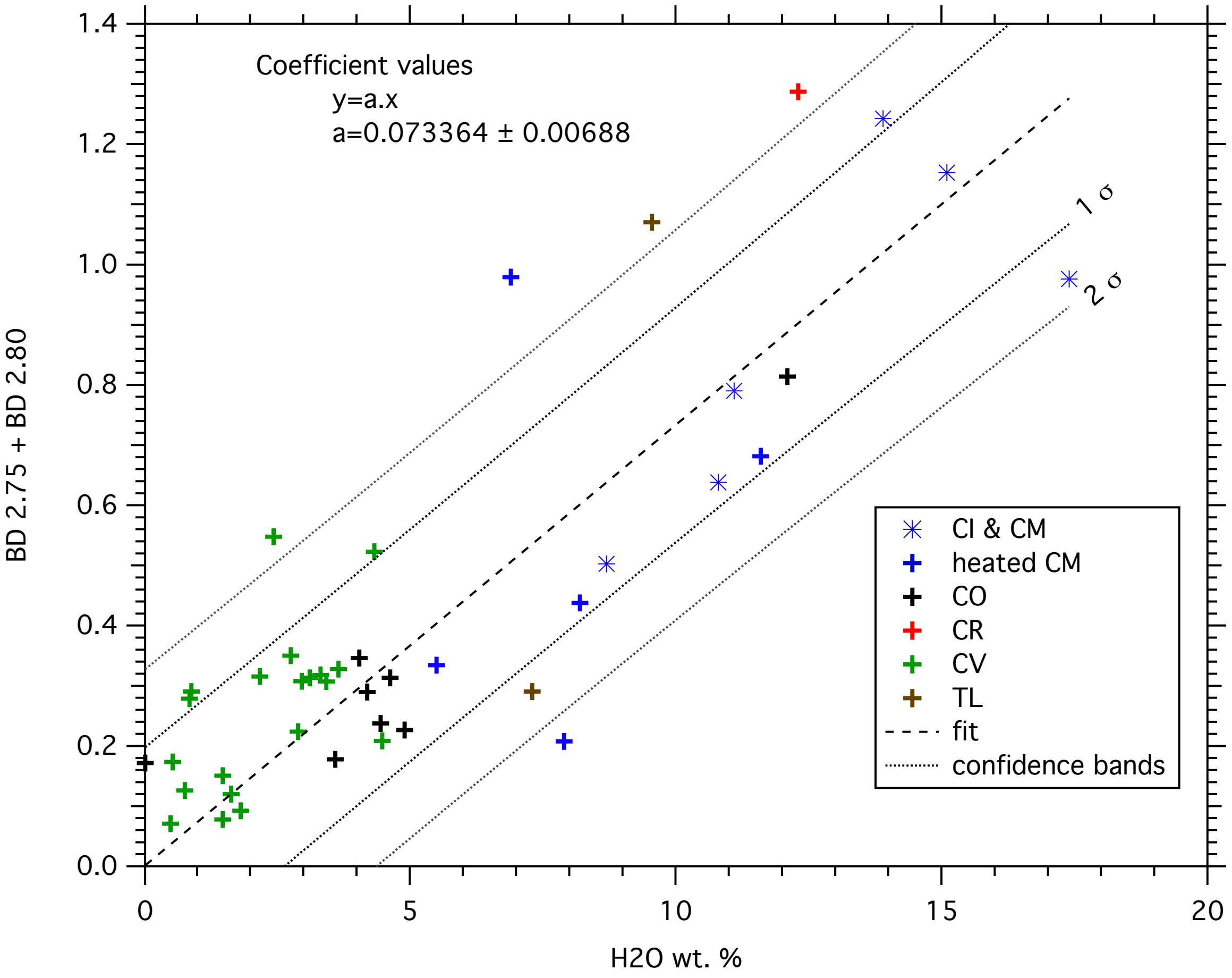

FIGURE 5: Spectral criterion used for the remote quantification of water compared to the water content (measured by TGA) of different classes of carbonaceous chondrites. W

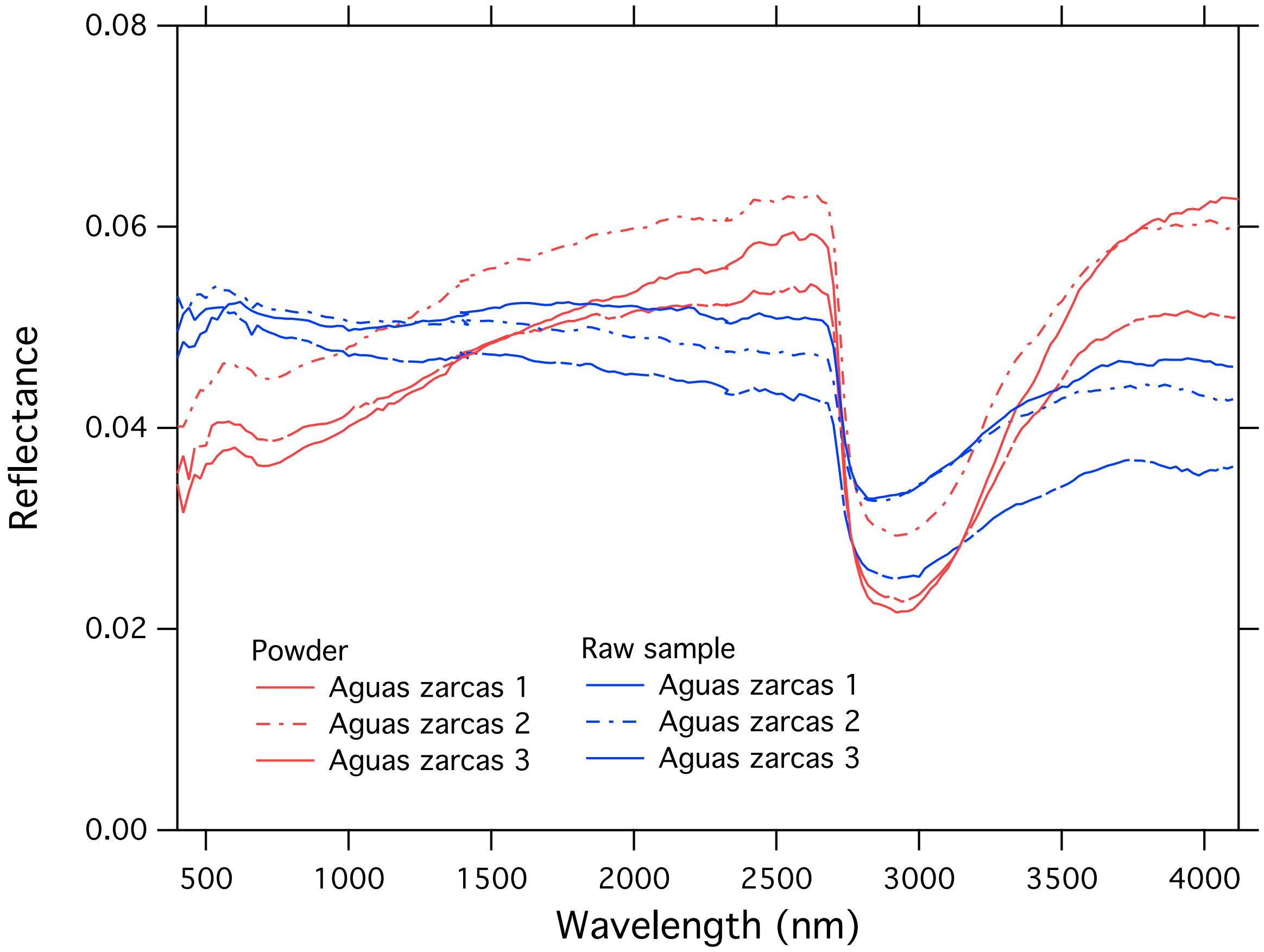

FIGURE 6: The effect of texture (rock vs powder) on the reflectance spectra of CM chondrites. Note that these spectra were not measured under vacuum. The three different samples correspond to three different lithologies from a large piece of the meteorite.

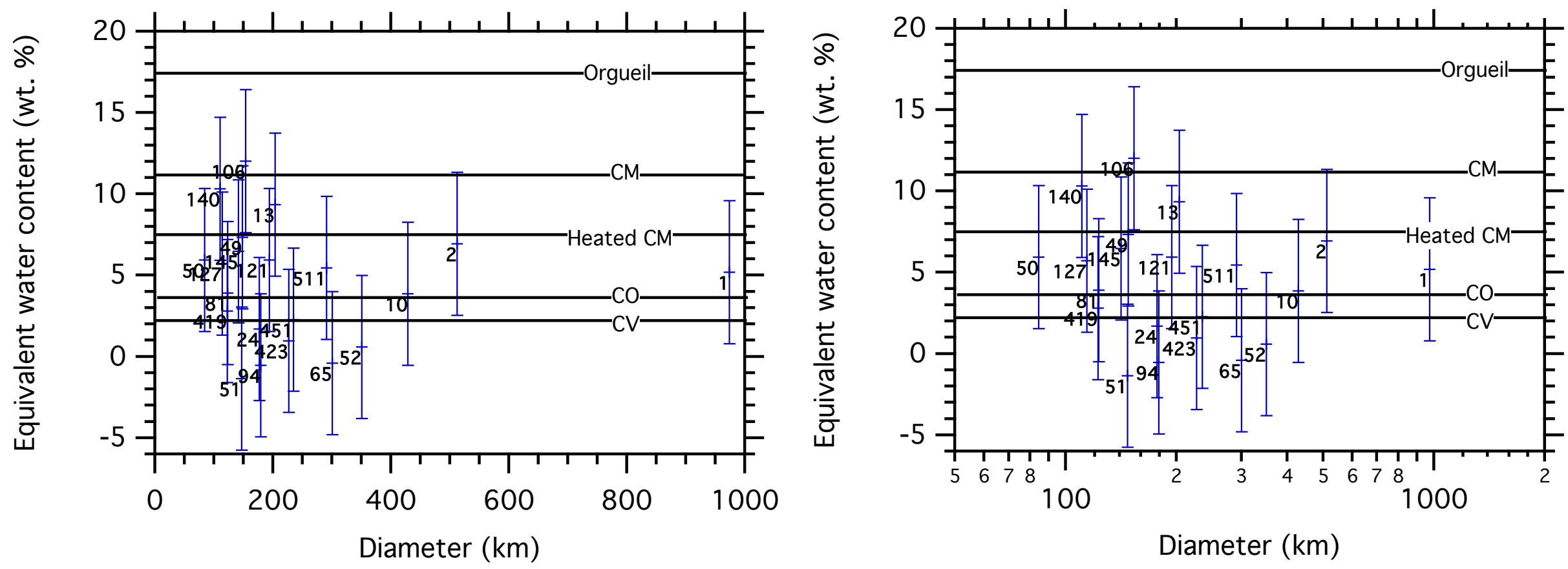

FIGURE 7: Equivalent water content estimated for large C-complex asteroids. The water content were estimated from reflectance spectra obtained from AKARI (Usui et al., 2018) with the exception of Ceres (Marchi et al., 2018). The horizontal lines are averaged for a given group, but note that there is important variability within each group (Table 1).

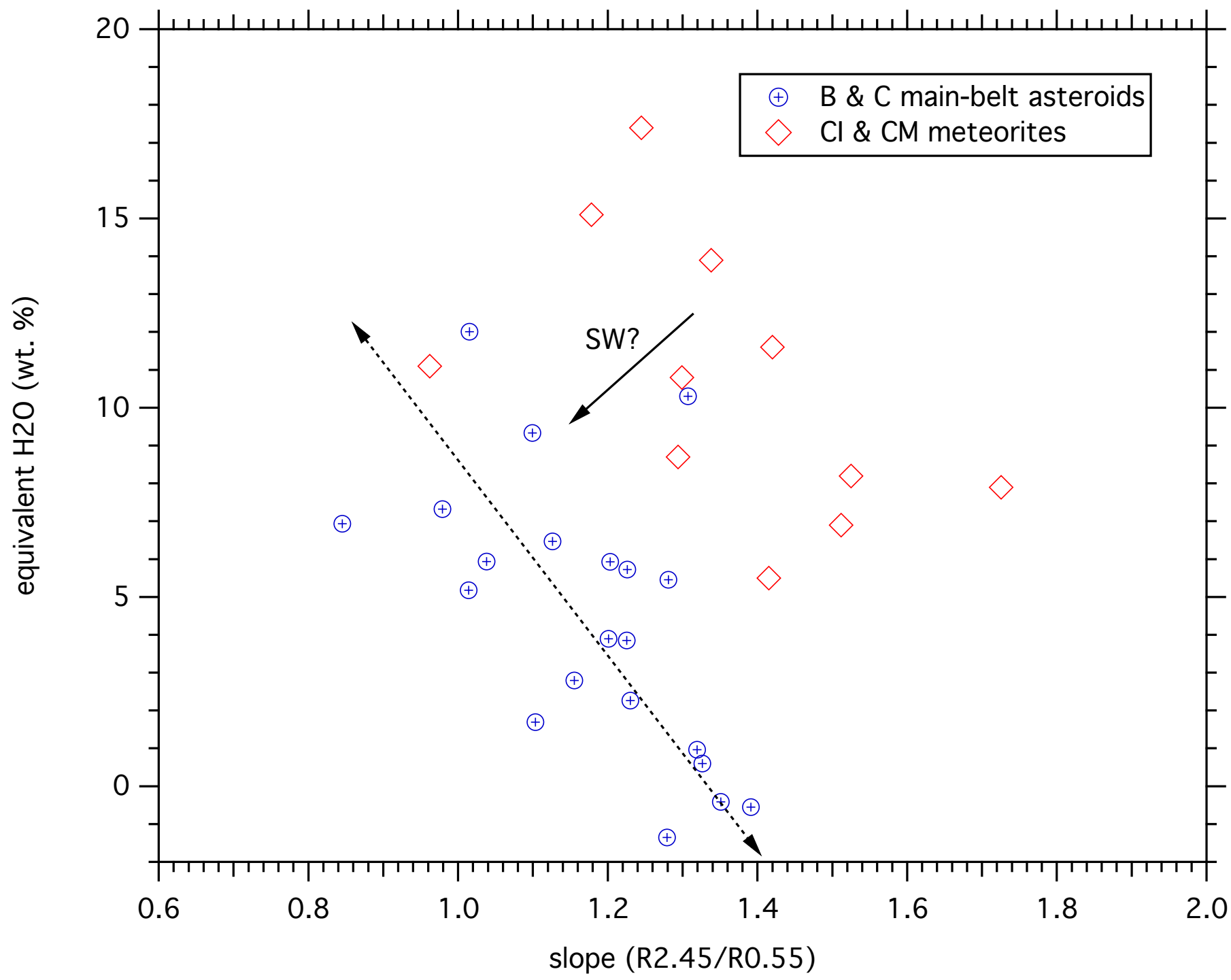

FIGURE 8: Vis-IR spectral slope as a function of equivalent water content C-complex main-belt asteroids and CI &CM meteorites. SW=Solar Wind.

Table 1: List of samples that were used in this study and the corresponding band depth at selected wavelength. The H2O estimated from TGA are taken from Garenne et al. (2014) and Bonal et al. (2020). Note that they correspond to the 200-770° mass loss as used in Garenne et al. for CI and CM and to the 200-900°C mass loss for CO, CV and CR (Bonal et al., 2020). *The star symbol denotes samples that were measured under vacuum at 25°C, unlike others that were measured under vacuum at 80-100°C. **The mass loss for Tagish Lake lithologies is from Gilmour et al. (2019)(200-700°C, average of two aliquotes). Paris lithologies were studied in Bonal et a. (2019) and Tagish Lake lithologies are described in Gilmour et a. (2020).

| Group | Name | | H2O from TGA | Band depth wavelength | | | | |
|---|---|---|---|---|---|---|---|---|
| | | | (wt. %) | 2750 | 2800 | 2750+2800 | 2900 | 3200 |
| CI | Orgueil | | 17.4 | 0.491 | 0.485 | 0.976 | 0.504 | 0.378 |
| CM | ALH 83100 | | 13.9 | 0.612 | 0.631 | 1.242 | 0.508 | 0.271 |
| CM heated | ALH 84033 | | 6.9 | 0.493 | 0.486 | 0.979 | 0.421 | 0.222 |
| CM | DOM 08003 | | 15.1 | 0.571 | 0.582 | 1.153 | 0.511 | 0.333 |
| CM heated | EET 96029 | | 8.2 | 0.255 | 0.182 | 0.438 | 0.28 | 0.148 |
| CM heated | MAC 88100 | | 11.6 | 0.297 | 0.385 | 0.682 | 0.198 | 0.074 |
| CM | MET 01070 | | 11.1 | 0.361 | 0.429 | 0.79 | 0.264 | 0.118 |
| CM heated | MIL 07700 | | 5.5 | 0.182 | 0.152 | 0.334 | 0.17 | 0.08 |
| CM | Murchison | | 10.8 | 0.361 | 0.277 | 0.638 | 0.325 | 0.212 |
| CM | QUE 97990 | | 8.7 | 0.308 | 0.195 | 0.503 | 0.303 | 0.175 |
| CM heated | WIS 91600 | | 7.9 | 0.105 | 0.103 | 0.208 | 0.07 | -0.004 |
| CM | Paris B1-4B | * | | 0.372 | 0.243 | 0.615 | 0.338 | 0.213 |
| CM | Paris B2-3D | * | | 0.051 | 0.038 | 0.089 | 0.032 | -0.003 |
| CM | Paros B2-3B | * | | 0.182 | 0.122 | 0.303 | 0.201 | 0.091 |
| CO | ALH 77003 | | 3.6 | 0.113 | 0.064 | 0.178 | 0.182 | 0.147 |
| CO | DOM 08006 | | 4.5 | 0.219 | 0.127 | 0.346 | 0.238 | 0.194 |
| CO | MIL 05024 | | 4.4 | 0.161 | 0.077 | 0.238 | 0.23 | 0.135 |
| CO | MIL 07193 | | 4.6 | 0.21 | 0.103 | 0.314 | 0.279 | 0.174 |
| CO | ALH 83108 | * | | 0.138 | 0.045 | 0.182 | 0.23 | 0.173 |
| CO | EET 092126 | * | | 0.155 | 0.045 | 0.2 | 0.298 | 0.234 |
| CO | MET 00737 | * | | 0.179 | 0.052 | 0.231 | 0.306 | 0.204 |
| CO | MIL 05104 | * | 4.2 | 0.193 | 0.097 | 0.29 | 0.293 | 0.219 |
| CO | MIL 07687 | * | 12.1 | 0.539 | 0.275 | 0.814 | 0.632 | 0.504 |
| CO | MIL 07709 | * | | 0.267 | 0.091 | 0.358 | 0.387 | 0.269 |
| CO | MIL 07709 | * | <0.1 | 0.112 | 0.059 | 0.172 | 0.142 | 0.11 |
| CO | Moss | * | | 0.067 | 0.023 | 0.09 | 0.139 | 0.099 |
| CO | Kainsaz | * | | 0.282 | 0.11 | 0.392 | 0.383 | 0.274 |
| CO | LAP 031117 | * | | 0.278 | 0.119 | 0.397 | 0.422 | 0.32 |
| CO | QUE 97416 | * | | 0.164 | 0.064 | 0.228 | 0.268 | 0.219 |
| CO | ALH 85003 | | 4.9 | 0.168 | 0.059 | 0.227 | 0.259 | 0.18 |
| | | | | | | | | |
| CR | EET 92042 | | | 0.335 | 0.189 | 0.524 | 0.38 | 0.297 |
| CR | GRA 95229 | | | 0.377 | 0.2 | 0.577 | 0.443 | 0.374 |
| CR | GRO 95577 | | 12.3 | 0.615 | 0.673 | 1.288 | 0.515 | 0.35 |
| CR | LAP 04720 | | | 0.307 | 0.157 | 0.464 | 0.37 | 0.299 |
| CR | MIL 090657 | | | 0.369 | 0.221 | 0.59 | 0.448 | 0.36 |
| | | | | | | | | |
| CV | ALH 81033 | | | 0.093 | 0.023 | 0.115 | 0.176 | 0.127 |
| CV | ALH 85006 | | | 0.265 | 0.122 | 0.387 | 0.403 | 0.303 |
| CV | Allende | | 0.8 | 0.102 | 0.024 | 0.126 | 0.168 | 0.133 |
| CV | Axtell | | 2.4 | 0.339 | 0.209 | 0.548 | 0.415 | 0.395 |
| CV | Efremovka | | | 0.224 | 0.088 | 0.311 | 0.324 | 0.263 |
| CV | GRA 06101 | | 0.5 | 0.113 | 0.06 | 0.173 | 0.178 | 0.122 |
| CV | GRO 95652 | | 3.7 | 0.225 | 0.103 | 0.328 | 0.324 | 0.275 |
| CV | Grosnaja | | 2.9 | 0.157 | 0.067 | 0.224 | 0.251 | 0.19 |
| CV | Kaba | | 4.5 | 0.168 | 0.041 | 0.209 | 0.31 | 0.234 |
| CV | LAP 02206 | | 0.5 | 0.037 | 0.034 | 0.071 | 0.068 | 0.047 |
| CV | LAR 06317 | | 3.4 | 0.242 | 0.066 | 0.307 | 0.422 | 0.339 |
| CV | MCY 05219 | | 1.5 | 0.061 | 0.017 | 0.078 | 0.132 | 0.114 |
| CV | MET 00761 | | 2.2 | 0.226 | 0.089 | 0.316 | 0.354 | 0.29 |
| CV | MIL 07002 | | 0.9 | 0.212 | 0.078 | 0.291 | 0.323 | 0.216 |
| CV | MIL 07277 | | 1.5 | 0.11 | 0.04 | 0.151 | 0.17 | 0.125 |
| CV | MIL 07671 | | 0.9 | 0.174 | 0.105 | 0.279 | 0.219 | 0.138 |
| CV | MIL 091010 | | 3.3 | 0.222 | 0.096 | 0.318 | 0.292 | 0.185 |
| CV | Mokoia | | 1.8 | 0.051 | 0.042 | 0.092 | 0.053 | 0.03 |
| CV | QUE 94688 | | 4.3 | 0.358 | 0.165 | 0.523 | 0.471 | 0.374 |
| CV | RBT 04302 | | 2.8 | 0.234 | 0.116 | 0.35 | 0.314 | 0.222 |
| CV | Leoville | | 3.0 | 0.212 | 0.096 | 0.308 | 0.222 | 0.175 |
| CV | Vigarano | | 3.1 | 0.189 | 0.125 | 0.314 | 0.203 | 0.129 |
| CV | MET 01074 | | 1.6 | 0.065 | 0.055 | 0.12 | 0.059 | 0.039 |
| | | | | | | | | |
| Tagish Lake | TL 11h | ** | 9.6 | 0.538 | 0.533 | 1.070 | 0.541 | 0.500 |
| Tagish Lake | TL4 | ** | 7.3 | 0.143 | 0.147 | 0.290 | 0.118 | 0.049 |

| # | Name | D km | $p_v$ | Tax. (B&D) | Tax. (Bus) | 3-μm shape | BD Combo | $H_2O$ wt. % | slope | Ref |
|---|---|---|---|---|---|---|---|---|---|---|
| 1 | Ceres | 974 | 0.087 | C | C | Ceres-like | 0.39 | 5.18 | 1.01 | 1 |
| 2 | Pallas | 512 | 0.150 | B | - | sharp | 0.75 | 6.94 | 0.85 | 2 |
| 10 | Hygiea | 429 | 0.066 | C | - | Ceres-like | 0.15 | 3.86 | 1.23 | 2 |
| 13 | Egeria | 203 | 0.086 | Ch | - | sharp | 0.67 | 9.34 | 1.10 | 2 |
| 24 | Themis | 177 | 0.084 | C | C | rounded | 0.53 | 1.70 | 1.10 | 2 |
| 49 | Pales | 148 | 0.061 | Ch | - | sharp | 0.85 | 7.32 | 0.98 | 2 |
| 50 | Virginia | 84 | 0.050 | Ch | C | sharp | 0.43 | 5.94 | 1.04 | 2 |
| 51 | Nemausa | 147 | 0.094 | Cgh | - | sharp | 0.79 | 0.00 | 1.28 | 2 |
| 52 | Europa | 350 | 0.043 | C | - | Europa-like | -0.37 | 0.60 | 1.33 | 2 |
| 65 | Cybele | 301 | 0.044 | Xk | C | rounded | -0.18 | 0.00 | 1.35 | 2 |
| 81 | Terpsichore | 123 | 0.048 | C | - | sharp | 0.32 | 3.90 | 1.20 | 2 |
| 94 | Aurora | 179 | 0.053 | C | - | - | -0.10 | 0.00 | 1.39 | 2 |
| 106 | Dione | 153 | 0.084 | Cgh | Cgh | sharp | 0.80 | 12.01 | 1.02 | 2 |
| 121 | Hermione | 194 | 0.058 | Ch | - | sharp | 0.48 | 5.93 | 1.20 | 2 |
| 127 | Johanna | 114 | 0.065 | Ch | Ch | sharp | 0.46 | 5.73 | 1.23 | 2 |
| 140 | Siwa | 111 | 0.067 | Xc | Cb | - | -0.05 | 10.31 | 1.31 | 2 |
| 145 | Adeona | 141 | 0.050 | Ch | Ch | sharp | 0.40 | 6.47 | 1.13 | 2 |
| 419 | Aurelia | 122 | 0.051 | C | Cb | - | 0.12 | 2.79 | 1.16 | 2 |
| 423 | Diotima | 227 | 0.049 | C | - | sharp? | 0.23 | 0.97 | 1.32 | 2 |
| 451 | Patientia | 235 | 0.071 | C | Cb | Europa-like | -0.16 | 2.26 | 1.23 | 2 |
| 511 | Davida | 291 | 0.070 | C | X | sharp | 0.44 | 5.46 | 1.28 | 2 |

Table 2: List of asteroids used in this study. Albedo and band shape were compiled in Usui et al., 2019. 1=DeSanctis et al., 2015: 2=Usui et al., 2018. B&D=Bus -DeMeo taxonomy, B : Bus taxonomy